# A Proposed Artificial Neural Network based Approach for Molecules Bitter Prediction

Huynh Quoc Anh Bui[1] and Trong Hop Do[1] and Thanh Binh Nguyen[1]

[1] *University of Information Technology, Vietnam National University HCM City*
`{22520038.gm, hopdt, binhnt}@uit.edu.vn`

**Abstract.** In recent years, the development of Artificial Intelligence (AI) has offered the possibility to tackle many interdisciplinary problems, and the field of chemistry is not an exception. Drug analysis is crucial in drug discovery, playing an important role in human life. However, this task encounters many difficulties due to the wide range of computational chemistry methods. Drug analysis also involves a massive amount of work, including determining taste. Thus, applying deep learning to predict a molecule's bitterness is inevitable to accelerate innovation in drug analysis by reducing the time spent. This paper proposes an artificial neural network (ANN) based approach (EC-ANN) for the molecule's bitter prediction. Our approach took the SMILE (Simplified molecular-input line-entry system) string of a molecule as the input data for the prediction, and the 256-bit ECFP descriptor is the input vector for our network. It showed impressive results compared to state-of-the-art, with a higher performance on two out of three test sets according to the experiences on three popular test sets: Phyto-Dictionary, Unimi, and Bitter-new set [1]. For the Phyto-Dictionary test set, our model recorded 0.95 and 0.983 in F1-score and AUPR, respectively, depicted as the highest score in F1-score. For the Unimi test set, our model achieved 0.88 in F1-score and 0.88 in AUPR, which is roughly 12.3% higher than the peak of previous models [1, 2, 3, 4, 5].

## 1   Introduction

The bitter taste of molecules is a sensory response mediated by specialized receptors on the tongue. It is a protective mechanism against ingesting potentially harmful compounds. In humans, the perception of Bitter taste is mediated by 25 G-protein-coupled receptors known as T2Rs or TAS2Rs [6]. These receptors are not only expressed in the oral cavity and various other tissues, including the gastrointestinal tract, upper airways, and even the heart [7,8,9]. Besides, The perception of bitterness is handled by 25 hTAS2R receptors that belong to the same set of signaling proteins [10].

Bitterness is frequently considered as an unpleasant taste in human perception, which is an critical aspect of development in the fields of food science and drug discovery. In food science, bitterness can be both detrimental and beneficial depending on a certain situation. For some special foods and beverages, such as tea, coffee and dark chocolate, it is crucial to notice that bitter taste as one of the unique flavors used to enhance the consumer experience. In contrast, it is reported that bitter taste appearing in other foods such as vegetables and medical herbs can negatively affect customer acceptance. In drug discovery, a large proportion of drugs accompanied with the bitter ingredient and almost all of them have bitter taste, which is a massive challenge in the development of oral medications. In addition, the bitterness of drugs can also indicate their toxicity, which can be a safety concern for patients.

However, the process of determining the bitterness of a specific compound is complex and lengthy; it demands considerable effort, requires enormous work with many chemical calculation methods, and is a time-consuming task. In addition, the problem of chemical diversity also has a significant impact on each individual compound; if there exists a slight difference between bitterant and non-bitterant molecules, they complicate matters [11]. Therefore, AI models have recently become practical tools to reduce the cost and time in determining the bitter characteristics of compounds. Despite the innovation of recent AI models, the main obstacle is the problem of lack of available data, which is also a primary factor that prevents these AI approaches from improving accuracy performances.

Many kinds of research involve predicting the bitterness of molecules, and these papers account for a large proportion of taste-predicting tasks. There are a vast number of formats in the data for molecules. Furthermore, most of the databases [12,13,14] store their molecules in SMILES format due to their advantages (specifically Compact representation, Human-readable, Unique, and possible to encode stereochemistry) [15]. In this research, we focus on the solutions based on SMILES strings to classify molecules as bitter or not.

There are commonly two main types of descriptors applied to previous research: structural and physicochemical descriptors. Particularly, structural descriptors consist of three kinds of format: Topological descriptors, Geometric descriptors, and Fingerprints. Besides, the information on physicochemical descriptors includes Electronic, Thermodynamic, and Spectroscopic descriptors. For each kind of format, they are encoded with distinct types of information, for instance, Topological descriptors: These describe the connectivity of atoms in a molecule and include measures such as the number of atoms, bonds, rings, and branches in a molecule, as well as graph-based properties such as the molecular weight, degree, and centrality.

According to the survey on computational taste predictors [16], numerous studies of molecule bitter classification have supplied many overall analyses by using both structural descriptors and physicochemical descriptors or the combination of two types [1,2,17]. On the other hand, the other studies focus on analyzing a specific type of descriptor, such as fingerprint [4,5] or physicochemical descriptors [3]. About the models, the survey separated the prior studies into these subcategories of machine learning: linear-based models [2,3,5], tree-based models [1,2,4,5] as well as Distance-based Models [5]. In the latest research, they successfully built a complete system to classify whether a molecule is bitter or not with considerable accuracy. In more detail, the study of e-bitter [5] has shown that they obtained 0.94 in the figure of f1-score on the test set split from their dataset, 0.92 in the study of BitterSweet Forest [4], and 0.84 in the study of BitterSweet [2]. however, they encountered many difficulties in classifying the external test sets that BitterPredict established, such as Phyto-Dictionary, Unimi, and Bitter-New, the f1-score of the e-bitter system gained 0.93 and 0.71 in Phyto-Dictionary and Unimi test sets, which are a significant decrease compared to their own test sets.

These research articles have shown a comprehensive perspective on this topic, advancing our knowledge regarding the relationship between bitter taste and molecular characteristics and contributing predictive models. Moreover, it can improve by carefully compiling bitter molecules and employing an extensive range of molecular descriptors. However, a limit of available data is still a big obstacle.

In this research, our purpose is to propose a new method to address the problem of molecule bitterness prediction via building a model based on the structure of the artificial neural network as well as utilizing the SMILES input by extracting their features from raw data, transforming to the feature vectors in forms of ECFP descriptors. We trained the models on the augmented dataset generated from the original dataset of the previous study of BitterSweet [2]. We increased the number of rows in the dataset using the augmentation strategy with reduced duplications introduced by Maxsmi [18]. Our models are evaluated by the external test sets established by BitterPredict [1]. The results on the three test sets show better performance on Phyto-Dictionary and Unimi test sets compared to other studies. Chapter II will describe in detail our approach and related background knowledge. The experiment results are in Chapter III, and finally, We presented our conclusions in Chapter IV.

## 2    Our Proposed Approach (EC-ANN)

Although we have many different approaches to solving the problem of the molecular's bitter character prediction, most of the approaches implement a similar general processing flow shown in Figure 1. below.

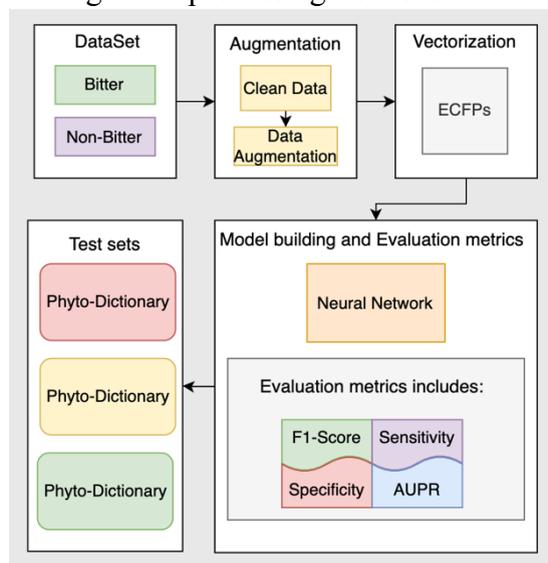

**Fig. 1.** Experimental procedure for the proposed Molecules Bitter Prediction model

This processing flow included five important stages, which are Data Preparation and Curation (preparing the dataset), Augmentation, Vectorization (feature generation), Model Building, and finally, Testing on external test sets. We aim to build the good molecule's bitterness classifier based on a well-curated dataset and some molecular descriptors. We aim to train our models on a combination of many verified resources from other research articles about the dataset. Continuously, all the invalid molecules appearing in the dataset must be removed and simultaneously conducted, the augmentation with a reduced duplicate strategy introduced by Maxsmi [18]. We implemented our ANN models utilizing an exhaustive set of molecular descriptors, including different bit lengths of ECFPs [19]. We evaluated our models based on the external test sets established by previous studies and took a comparison with other reports published recently [1,2,3,5].

## 2.1 Data Preparation and Curation.

First, We curated information on molecules with bitter-sweet taste from various resources ranging from scientific publications to books. After removing molecules for which exact information about their bitter taste was either unavailable or conflicting. We split the datasets into training and test sets such that the latter corresponded to the external validation/test sets established by BitterPredict [1] for Bitter/non-Bitter prediction. Our training set consists of about 1436 non-bitter molecules and 797 bitter molecules.

The external test sets which were established by the research of BitterPredict [1] are supposed to be the evaluation basis for our model performances. Specifically, the BitterPredict [1] has provided three test sets: Phyto-Dictionary, Unimi, and Bitter-new. The number of molecules in each data set was shown in Table I.

**Table 1.** Our training set and test sets

|  | Bitter | Non-Bitter | Total |
|---|---|---|---|
| Our training data | 797 | 1436 | 2233 |
| Phyto-Dictionary test set | 55 | 33 | 88 |
| Unimi test set | 23 | 33 | 56 |
| Bitter-New set | 27 | 0 | z27 |

## 2.2 Augmentation

In fact, there is plenty of available molecular data from various resources. However, they exist in unidentified taste forms, and just several of them are verified as bitterant ones. This situation hinders applying Deep Learning, which is truly greedy and demands a massive quantity of data. Thus, the augmentation of the smiles string is a prerequisite step that plays an important role in improving the models' performance.

The augmentation of SMILES (Simplified Molecular Input Line Entry System) has proven as a crucial process to develop robust machine learning models in the task of molecular properties prediction. However, this process has to be taken into explicit consideration due to the risk of a high degree of redundancy and similarity between the original and augmented SMILES string. This is where the study of Maxsmi [18] comes in; they have introduced a method for SMILES augmentation with reduced duplication.

The study of Maxsimi [18] has developed a Python library for SMILES augmentation, which is based on two main techniques. In the first stage, they perform a randomization technique that uses a SMILES string as an input, randomly generates a lot of new SMILES strings, and alters them to the original one. Then, the filter technique is taken to reject many augmented SMILES strings, ensuring the uniqueness of augmentation. In more detail, Maxsmi [18] incorporates a unique fingerprint-based similarity measure, which guarantees an acceptable distance to the original string, also known as ensuring chemical diversity. Additionally, Maxsmi leverages the RDKit library to ensure the chemical feasibility in each generated SMILES string. They have successfully provided an effective solution for SMILES augmentation, which not only improves the machine learning model performance but also reduces the redundancy between SMILES strings.

The study of Maxsmi [18] provides an effective solution to the issue of SMILES augmentation with reduced duplication. By utilizing a unique fingerprint-based similarity measure and the RDKit library, Maxsmi [18] ensures that the generated SMILES strings are chemically diverse and feasible. This technique improves the performance of machine learning models and saves time and computational resources by reducing redundancy and similarity between SMILES strings.

As mentioned previously, deep learning being data greedy and both Physicochemical and bioactivity databases being meager, elaborate techniques must be integrated to unleash the full potential of deep neural networks. In this context, data augmentation, in general, and more specifically SMILES augmentation, is a powerful assistance in molecular prediction. From a machine-learning perspective, data augmentation allows the model to see the same object from different angles. It has been successfully applied in image classification, where images transform such as flipping, coloring, cropping, rotating, and translating. From a computational perspective, SMILES augmentation is advantageous because generating ran- dom SMILES is fast and memory efficient, and even though training a model may be more computationally expensive, it remains cheap to evaluate.

In this research, we used the augmentation technique to reduce the duplicates introduced by Maxsmi [18]. Removing duplicate entries is common in data wrangling. In the context of SMILES augmentation, this translates to discarding duplicates after generating the random SMILES. For data set $D$, the final number of data points after augmentation varies according to the augmentation number, i.e. the number of times a sample is drawn from the valid SMILES space and the size of the molecules in the data set. We built the larger dataset from 2233 rows to a massive dataset with 22465 rows with 22465 non-bitter molecules and 16267 bitterants. Due to the disparity between non-bitter and bitter molecules, we leverage augmentation by increasing the number of bitterants with 20 samples for each molecule and 15 for each non-bitter molecule. It lowers the ratio between non-bitter molecules and bitterant, which decreases from 1.80 to 1.38. The former corresponds to the original dataset, and the latter is for the augmented one.

## 2.3 ECFPs and Vectorization

ECFPs (Extended Connectivity Fingerprints) are widely known as circular topological fingerprints, which encode the topology of a molecule. They were designed for molecular characterization, similarity searching, and structure-activity modeling based on a circular substructure that grows iteratively from each atom in a molecule. By determining the particular radius, they extend from an atom with the radius of the substructure by one bone in each iteration. Finally, the algorithm results in an ECFP substructure in a vector of bits; each bit in that vector corresponds to the presence or absence of a partiture within the molecule. Specifically, the formation of ECFPs descriptors can be understood by this process. The molecule can be represented as a graph format with nodes and edges, the former correspond to atoms and the latter represents the bounds in molecules. The ECFP algorithm firstly performs a breadth-first manner, iteratively traversing the graph and adding new nodes to the structure within the defined radius. Subsequently, the ECFP substructure maps into a particular bit in the vector through a specific hash function. This process is repeated recursively until all the atoms in the predefined radius have been visited.

In general, ECFPs as molecular fingerprints have provided various advantages. They encode important topological information about the molecule and are also computationally efficient, making them well-suited for large-scale virtual screening experiments. For illustration, topological information contains information such as the presence of cycles and branching patterns, which are effectively used to classify their biological activities.

In our research, We propose an application based on ECFP, which forms a 256-bit vector fed to the models. This process can be handled easily with support from the RD-kit library. The first step is procuring the Mol objects from our SMILES strings, continuously followed by creating a Morgan Fingerprint from the preceding Mols with these parameters: radius of two, 256-bit vector.

### 2.4  Model Architecture

Our research emphasizes the application of deep learning. It proposes the model of artificial neural networks to process the information conveyed from the descriptors to tackle the problem of molecules' bitterness prediction. The specification of our model is shown in Figure 2 below.

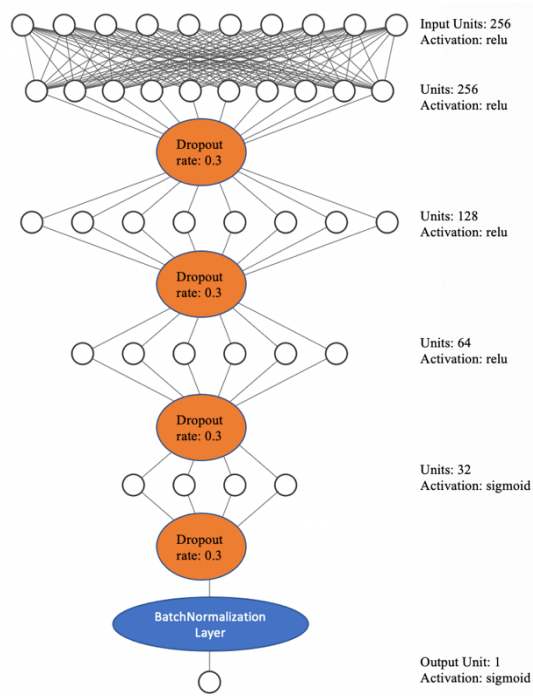

**Fig. 2.** The structure of the proposed artificial neural network (EC-ANN)

In the proposed EC-ANN model, we use 256 units in the input layer. The next four hidden layers contain 256, 128, 64, and 32 units. The dropout technique has been used between each input layer, hidden layer, and output layer. The batch normalization technique was applied before the output layer. The learning rate parameter is set to 0.00025, and the optimiser used is Adam which is a stochastic gradient descent method that is based on adaptive estimation of first-order and second-order moments. Finally, the binary cross entropy is chosen as a loss function for our model.

**Table 2.** Model performances on training set

| Model/ Descriptor | F1-score | AUPR |
|---|---|---|
| EC-ANN - 256-bit ECFP | 0.988 | 0.999 |

## 3    Our Experiment Results

We trained our models using the dataset BitterSweet provided. Each dataset row contains SMILES strings, Canonical SMILES strings, and binary data, which defines a bitter compound as True or False. As an illustration, a row can be OCC(C(C(CO)O)O)O, OCC(C(C(CO)O)O)O, False, which are SMILES string, Canonical SMILES string, and Bitter identification, respectively. Our EC-ANN Network will use the set of descriptors as its input, including 256-bit ECFP [19], to classify the bitterness of molecules into False or True. In this section, we will consider our model based on external test sets, which are Phyto-Dictionary, Unimi, and Bitter-new and have a comprehensive comparison with other previous SOTA (state-of-the-art) models [8,7,12,16].

|  | Molecular Descriptors | Model | Phyto-Dictionary | | | | UNIMI | | | | Bitter-New | | | |
| --- | --- | --- | --- | --- | --- | --- | --- | --- | --- | --- | --- | --- | --- | --- |
|  |  |  | Sn | Sp | F1 | AUPR | Sn | Sp | F1 | AUPR | Sn | Sp | F1 | AUPR |
| BitterSweet | Canvas | RF-Boruta | 0.980 | 0.76 | 0.932 | 0.959 | 0.826 | 0.727 | 0.745 | 0.810 | 0.652 | — | 0.789 | — |
|  | Dragon2D | RLR-PCA | 0.979 | 0.64 | 0.904 | 0.970 | 0.609 | 0.970 | 0.737 | 0.847 | 0.609 | — | 0.757 | — |
|  | Dragon2D/3D | RF-PCA | 0.980 | 0.68 | 0.914 | 0.977 | 0.913 | 0.636 | 0.750 | 0.746 | 0.913 | — | 0.955 | — |
|  | ECFPs | AB | 0.959 | 0.84 | 0.940 | 0.970 | 0.913 | 0.697 | 0.778 | 0.783 | 0.652 | — | 0.789 | — |
| BitterX | Handbook of Molecular Descriptors | SVM | 0.939 | 0.308 | 0.814 | — | 0.652 | 0.562 | 0.577 | — | 0.739 | — | 0.850 | — |
| BitterPredict | Canvas | AB | 0.980 | 0.692 | 0.914 | — | 0.783 | 0.848 | 0.783 | — | 0.739 | — | 0.850 | — |
| e-Bitter | ECFPs | CM01 | 0.980 | 0.769 | 0.932 | — | 0.913 | 0.545 | 0.712 | — | 1.000 | — | 1.000 | — |
| Our proposed Model | ECFPs | EC-ANN | 0.927 | 0.969 | 0.953 | 0.983 | 0.956 | 0.848 | 0.88 | 0.880 | 0.680 | — | 0.784 | — |

**Fig. 4.** Our results and previous results on the external test sets: Phyto-Dictionary, Unimi, Bitter-New.

For the Bitter prediction task, we analyzed our results with other models from BitterX, BitterPredict, e-bitter, and BitterSweet based on independent external test sets.

In the Phyto-Dictionary test set, our proposed model (EC-ANN) resulted in 0.927, 0.969, 0.953, and 0.983 in sensitivity, specificity, F1-score, and the Average Under Precision-Recall Curve (AUPR), respectively. As an observation from the above results, our model has outperformed other models from BitterX, BitterPredict, e-bitter, and BitterSweet with attained disparities of 17.0%, 4.2%, 2.2%, and 1.3% in the F1 score, respectively. Additionally, the model has witnessed an impressive result with 0.983 for the AUPR score, which strongly affirms that our model performs almost perfectly on this test set. Besides, our model recorded a 0.969 score in the figure of specificity, which is a peak of all the experiments and higher than the previous highest point.

In the Unimi test set, our model encounters many difficulties distinguishing between the molecules as bitterants. Much of that comes from molecules with similar scaffolds but different tastes. However, our Neural Network EC-ANN based on ECFP achieved a better performance in F1-score and AUPR score: 0.88 and 0.88 compared to other models from other studies. In more detail, our EC-ANN based on the ECFP descriptor model has a higher score with disparities of approximately 13.1%, 52.5%, 12.3%, and 23.5% compared to the results obtained from BitterSweet, BitterX, BitterPredict, and e-Bitter respectively. Additionally, our model's sensitivity and specificity archived high scores, namely 0.956 and 0.848 in sensitivity and specificity scores. The results determined that our model sensitivity, AUPR, and F1-score are the highest recorded score in the Unimi test set compared to other previous State-of-the-art models.

The Bitter-new molecule set was the most miniature bitter/non-bitter test set, comprising 27 bitter molecules. The lack of presence of non-bitter molecules made the calculation of all metrics besides sensitivity infeasible. Our Model obtained was 0.680 in the sensitivity score and 0.784 in the F1 score.

## 4 Conclusion

In this research, we proposed an artificial neural network-based approach for predicting the bitterness of molecules. To achieve this, we utilized augmentation with a reduced duplication strategy to generate a larger dataset from the original data. Our models took SMILES strings as input, from which we extracted ECFP descriptors. Our structure-taste relationship models used these descriptors to learn the chemical characteristics of each molecule and predict its bitterness. To evaluate the performance of our models, we used external test sets - Phyto-Dictionary, Unimi, and Bitter-new. These test sets are crucial for rating the performance of our models and providing a comprehensive comparison with other previous models. Our results showed the Phyto-Dictionary test set. Our EC-ANN model achieved the highest AUPR of 0.983 and F1-score of 0.953. Moving on to the Unimi test set, it achieved an F1-score of 0.88 and an AUPR of 0.88, which is 12.3% higher than the peak of previous models.

## 5 Acknowledgment

This research received support from the University of Information Technology, Vietnam National University of Ho Chi Minh City. We thank our colleagues from the Information System Faculty who provided insight and expertise that greatly assisted the research.